\documentclass[usenatbib]{pasj01} 

\Received{2021/10/12}
\Accepted{2022/03/02}
 
\usepackage{hyphenat} 
\usepackage[modulo]{lineno}
\usepackage{url}
\usepackage{comment}
\usepackage{upgreek}
\usepackage{diagbox}
\makeatletter
\def\@fnsymbol#1{\ensuremath{\ifcase#1\or \dagger\or \ddagger\or
   \mathsection\or \mathparagraph\or \|\or **\or \dagger\dagger
   \or \ddagger\ddagger \else\@ctrerr\fi}}
    \makeatother

\begin{document} 

\title{ 
Passive spiral galaxies deeply captured by Subaru Hyper Suprime-Cam}

\author{Rhythm \textsc{Shimakawa}\altaffilmark{1}$^\ast$\thanks{NAOJ Fellow}}
\altaffiltext{1}{National Astronomical Observatory of Japan (NAOJ), National Institutes of Natural Sciences, Osawa, Mitaka, Tokyo 181-8588, Japan}
\email{rhythm.shimakawa@nao.ac.jp (RS)}

\author{Masayuki \textsc{Tanaka}\altaffilmark{1}}

\author{Connor \textsc{Bottrell}\altaffilmark{2}}
\altaffiltext{2}{Kavli Institute for the Physics and Mathematics of the Universe (WPI), UTIAS, University of Tokyo, Kashiwa, Chiba 277-8583, Japan}

\author{Po-Feng \textsc{Wu}\altaffilmark{3}}
\altaffiltext{3}{Academia Sinica Institute of Astronomy and Astrophysics, Taipei 10617, Taiwan}

\author{Yu-Yen \textsc{Chang}\altaffilmark{3,4}}
\altaffiltext{4}{Department of Physics, National Chung Hsing University, 40227, Taichung, Taiwan}

\author{Yoshiki \textsc{Toba}\altaffilmark{3,5,6}}
\altaffiltext{5}{Department of Astronomy, Kyoto University, Kitashirakawa-Oiwake-cho, Sakyo-ku, Kyoto 606-8502, Japan}
\altaffiltext{6}{Research Center for Space and Cosmic Evolution, Ehime University, 2-5 Bunkyo-cho, Matsuyama, Ehime 790-8577, Japan}

\author{Sadman \textsc{Ali}\altaffilmark{1}}



\maketitle

\begin{abstract}
This paper presents a thousand passive spiral galaxy samples at $z=$ 0.01--0.3 based on a combined analysis of the Third Public Data Release of the Hyper Suprime-Cam Subaru Strategic Program (HSC-SSP PDR3) and the GALEX--SDSS--WISE Legacy Catalog (GSWLC-2). 
Among 54871 $gri$ galaxy cutouts taken from the HSC-SSP PDR3 over 1072 deg$^2$, we conducted a search with deep-learning morphological classification for candidates of passive spirals below the star-forming main sequence derived by UV to mid-IR SED fitting in the GSWLC-2. 
We then classified the candidates into 1100 passive spirals and 1141 secondary samples based on visual inspections. 
Most of the latter cases are considered to be passive ringed S0 or pseudo-ringed galaxies. 
The remainder of these secondary samples has ambiguous morphologies, including two peculiar objects with diamond-shaped stellar wings. 
The selected passive spirals have a similar distribution to the general quiescent galaxies on the EW$_\mathrm{H\delta}$--D$_n$4000 diagram and concentration indices.
Moreover, we detected an enhanced passive fraction of spiral galaxies in X-ray clusters. 
Passive spirals in galaxy clusters are preferentially located in the midterm or late infall phase on the phase-space diagram, supporting the ram pressure scenario, which has been widely advocated in previous studies. 

The source catalog and $gri$-composite images are available on the HSC-SSP PDR3 website (\url{https://hsc.mtk.nao.ac.jp/ssp/data-release/}). 
Future updates, including integration with a citizen science project dedicated to the HSC data, will achieve more effective and comprehensive classifications. 
\end{abstract}


\section{Introduction}

To understand the fate of spiral galaxies, including our own Galaxy, passive spiral galaxies (termed `anemic' spirals; \cite{vandenBergh1976}) yield important insights into their transition processes from the star-forming to the quiescent phase. 
They would also uniquely give a hint at the formation and destruction mechanisms of spiral structures (see theoretical studies, e.g. \cite{Lin1964,Goldreich1965,Sellwood1984,Bertin1989,D'Onghia2013,Dobbs2014}). 

From a historical perspective, since the discovery of the significant presence of passive spirals in rich clusters \citep{vandenBergh1976}, the physical origins of passive spirals have been widely linked to environmental phenomena \citep{Goto2003,Yamauchi2004,Koopmann2004,Boselli2006,Moran2007,Blanton2009,Masters2010}. 
Additionally, cold gas observations have reported that passive spiral galaxies are H{\sc i}-deficient but not in molecules compared to normal star-forming spirals, suggesting a lack of persist gas supply owing to truncation of the outer H{\sc i} disk, which may originate from environmental effects in some cases (e.g. \cite{Bothun1980,Giovanelli1983,Cayatte1994,Bravo-Alfaro2001,Elmegreen2002}). 
Ram pressure stripping in clusters or falling subclusters (referred to as pre-processing) can lead to strangulation and slow star formation quenching \citep{Gunn1972,Fujita2004,Kawata2008,Hamabata2019,Rhee2020}. 

However, a consensus has yet to be reached on these matters, as their abundance is overwhelmingly small (e.g., $\sim6$\% in \cite{Masters2010}, and see also references therein). 
The Sloan Digital Sky Survey (SDSS; \cite{York2000}) has played a key role in sample collection over the last two decades. 
In addition, UV and IR photometric data, such as the Galaxy Evolution Explorer (GALEX; \cite{Martin2005}) and the Wide-field Infrared Survey Explorer (WISE; \cite{Wright2010}) are crucial to make a reliable selection \citep{Fraser-McKelvie2016}. 
Conversely, much deeper and sharper images with larger aperture telescopes are highly desired to conduct an in-depth passive spiral search, which can cover faint spiral structures missing in previous surveys and also resolve pseudo-rings with smaller pitch angles of spiral arms. 
Such objects with less-prominent spiral features provide a less biased insight into the late evolutionary stage of spiral galaxies. 

The Hyper Suprime-Cam Subaru Strategic Program (HSC-SSP; \cite{Aihara2018,Miyazaki2018,Furusawa2018,Kawanomoto2018,Komiyama2018}) on the 8.2 m Subaru Telescope satisfies these stringent requirements. 
The HSC-SSP surveys 1400 deg$^2$ areas in $grizy$ broadbands approximately 3 mag deeper and with $\sim2$ times better seeing size compared to the SDSS. 
With such high-quality data, we can expect to capture fainter spiral arms not detected in previous surveys and those with smaller pitch angles mistaken for ringed galaxies. 
Many studies have demonstrated the utility of the HSC-SSP data, such as galaxy morphological classifications \citep{Tadaki2020,Martin2020,Shimakawa2021}, weak lensing measurements \citep{Oguri2018,Oguri2021}, and anomaly detection \citep{Storey-Fisher2021,Tanaka2021}. 
Therefore, the HSC-SSP data will enable us to \emph{significantly} expand upon existing passive spiral samples, offering a unique opportunity to understand the transition phase from spiral galaxies to early-type galaxies statistically. 

With this motivation, this study aims to establish a large sample of passive spiral galaxies by using the HSC-SSP PDR3 \citep{Aihara2021}. 
The remainder of this paper is organized as follows. 
We first overview our dataset and deep learning-based methodology to select spiral galaxies at $z=0.01{\rm -}0.3$ (section~\ref{s2}).
Subsequently, we choose passive spiral galaxies based on the GALEX--SDSS--WISE Legacy Catalog (GSWLC-2; \cite{Salim2018}) and describe our passive spiral catalog (Section~\ref{s3}). 
In section~\ref{s4}, we compare the spectral properties of the passive spirals with those of the entire sample, and investigate their dynamic associations with massive galaxy clusters for the cluster members. 
Finally, we summarize our results and prospects (section~\ref{s5}). 
Following the GSWLC-2 catalog used in this study \citep{Salim2018}, the AB magnitude system \citep{Oke1983} and a \citet{Chabrier2003} stellar initial mass function (IMF) are adopted. 
Moreover, we assume a flat WMAP7 cosmology ($H_0=70$ km~s$^{-1}$Mpc$^{-1}$, $\Omega_m=0.27$; \cite{Komatsu2011}). 

\section{Data and methodology}\label{s2}

In the subsections that follow, we introduce the salient features of the HSC-SSP and the selection of 54871 main targets at $z=$ 0.01--0.3 (section~\ref{s21}). 
Subsequently, we describe the color-image construction (section~\ref{s22}), and the calibration of our deep learning classifier of spiral morphologies (section~\ref{s23}).

\subsection{Sample selection}\label{s21}

The HSC-SSP PDR3 provides a science-ready catalog and coadd imaging data, which were taken through 278 nights and reduced by the dedicated pipeline ({\tt hscPipe version 8}; \cite{Bosch2018}). 
The survey field is split on the database into areas of approximately $1.7\times1.7$ deg$^2$, termed {\tt tracts}, and further divided into approximately $12\times12$ arcmin$^2$ regions called {\tt patches} \citep{Aihara2021}. 
To exclude poor seeing data, we employ only patch regions with seeing FWHM (full-width-half-maximum) smaller than 1.0 arcsec in all $gri$ bands, which amounts to the total survey area of 1072 deg$^2$. 

As a starting point, we use 63504 bright galaxies at $z=$ 0.01--0.3 in the HSC-SSP PDR3 Wide layer\footnote{\url{https://hsc.mtk.nao.ac.jp/ssp/data-release/}} \citep{Aihara2021}, all of which have spec-$z$ counterparts in the SDSS DR15 \citep{Blanton2017,Aguado2019,Bolton2012} and the GALEX--SDSS--WISE Legacy Catalog (GSWLC-2; \cite{Salim2016,Salim2018}). 
The targets are mostly based on the SDSS Main Galaxy Sample ($r_\mathrm{petro}<17.77$; \cite{Strauss2002}) as they select bright SDSS sources above the \citet{Petrosian1976} magnitude limit, $r_\mathrm{petro}<18$. 
The GSWLC-2 has derived their physical properties, including stellar masses and star formation rates (SFRs) with UV to mid-IR spectral energy distribution (SED) fitting using CIGALE \citep{Noll2009} with stellar population synthesis models in \citet{Bruzual2003}, based on GALEX GR6/7 \citep{Martin2005}, SDSS DR7 \citep{Abazajian2009}, and 22 $\mu$m data from ALLWISE source catalog \citep{Lang2016}. 
Based on these datasets, the GSWLC-2 provided three separate catalogs depending on the depth of the UV survey (all-sky, A; medium, M; and deep, D). The GSWLC-A catalog is employed in this study to homogeneously select passive spirals across the survey field, unless otherwise noted. 

All 63504 sources are detected in the $grizy$ broadband data of the HSC-SSP PDR3 and are not influenced by nearby bright stars ($G<18$ mag; \cite{Aihara2021}), cosmic rays, or bad pixels, by applying the following criteria in the {\tt SQL} query:\\
{\tt forced.isprimary=True},\\
{\tt inputcount\_flag\_noinputs=False},\\
{\tt sdssshape\_flag\_badcentroid=False},\\
{\tt pixelflags\_edge=False},\\ 
{\tt pixelflags\_crcenter=False},\\ 
{\tt pixelflags\_bad=False},\\
{\tt psfflux\_flag=False},\\
{\tt mask\_brightstar\_halo=False},\\ 
{\tt mask\_brightstar\_ghost=False},\\
{\tt mask\_brightstar\_blooming=False}.\\
For details on these catalog flags, refer to \citet{Coupon2018,Aihara2021,Bosch2018}. 
In addition, point sources are removed based on the threshold of {\tt i\_psfflux\_mag$-$i\_cmodel\_mag$>$0.2} (see also \cite{Strauss2002,Baldry2010}). 
Consequently, we are left with 54871 galaxies at $z=$ 0.01--0.3 over 1072 deg$^2$.

\subsection{Image construction}\label{s22}

For the selected targets, we constructed RGB color cutouts in the {\tt png} format with a size of $(4\times\mathrm{R90})^2$ arcsec$^2$, where R90 is a radius within which 90\% of the Petrosian aperture flux in the $r$-band is included. 
We adopted the R90 values of our targets from the SDSS DR15 ({\tt petroR90}; \cite{Aguado2019}) by cross-matching between our sample and the SDSS sources within a radius of 1 arcsec. 
Flux scaling generally follows our previous work \citep{Shimakawa2021}, which adopts an inverse hyperbolic sine function, arcsinh \citep{Lupton1999,Lupton2004}. 
The arcsinh normalization has been widely employed in previous studies related to galaxy morphology classification, such as the {\tt Galaxy Zoo} project \citep{Lintott2008,Lintott2011,Willett2013}. 
Because the arcsinh stretch has also been used in the citizen astronomy program based on the HSC imaging data ({\tt GALAXY CRUISE}\footnote{\url{https://galaxycruise.mtk.nao.ac.jp}}), we can integrate this analysis with a large training dataset provided by this citizen science program in the near future. 

After homogenizing the seeing FWHM of the images to 1.0 arcsec by the Gaussian smoothing, we perform the arcsinh normalization with a code provided by the HSC-SSP pipeline team\footnote{\url{https://hsc-gitlab.mtk.nao.ac.jp/ssp-software/data-access-tools/tree/master/pdr3/colorPostage}}. 
The normalized flux $f(x)$ can be described as follows:
\begin{equation}
    f(x) = \frac{x}{I}\cdot\frac{\mathrm{arcsinh}(\alpha I)}{\mathrm{arcsinh}(\alpha)} + \beta,
\label{eq1}
\end{equation}
where $I={\displaystyle \Sigma_{x\in\{gri\}}}/3$ and $x$ is the pixel value ($f_\nu$) on coadd image in each filter.
Standardization by the mean flux over the $gri$-band photometry ($I$) allows us to minimize color saturation. 
We adopted default values of the scaling parameter of $\alpha=\exp(10)$ and the bias parameter of $\beta=0.05$. 
In addition, we set the lower and upper limits to 0 and 1, respectively. 
Examples of the resultant color images are shown in figures~\ref{fig4}, \ref{fig5}, and \ref{fig7}.

\subsection{Deep learning model}\label{s23}

\begin{figure}
\begin{center}
\includegraphics[width=7.5cm]{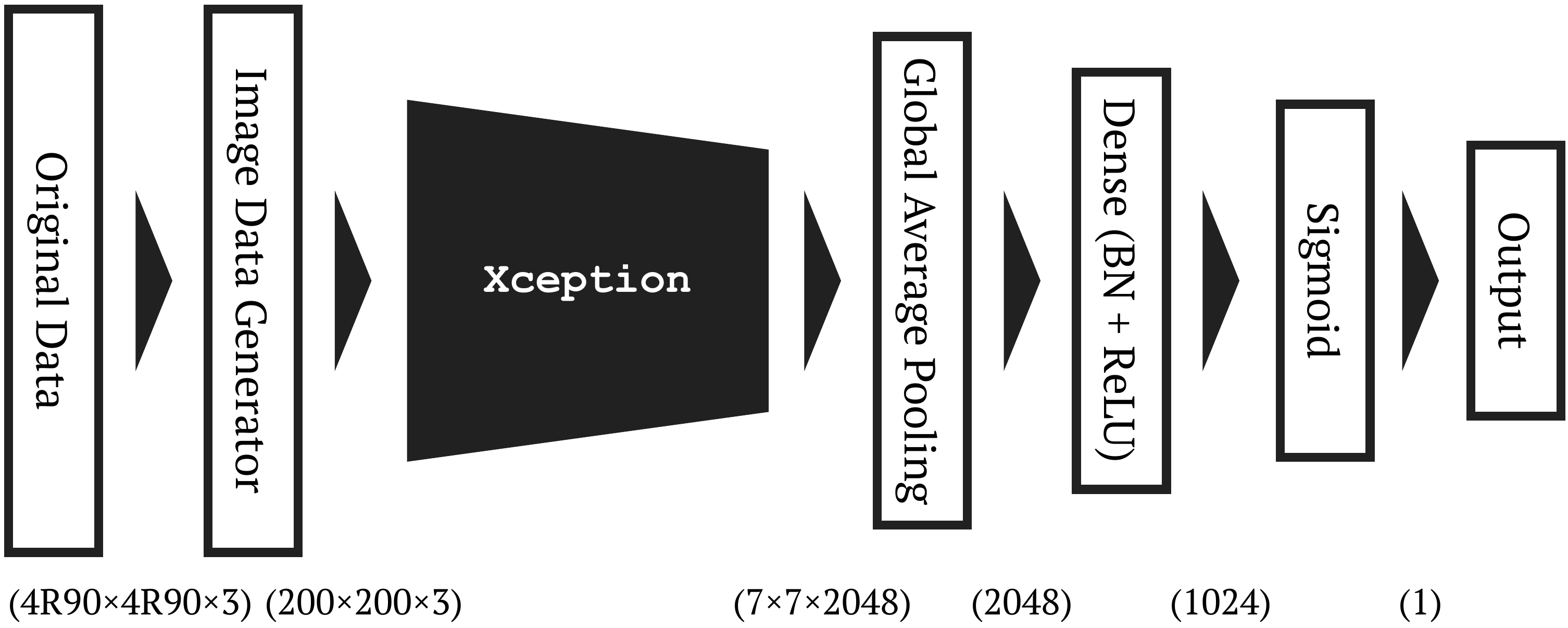}
\end{center}
\caption{
An outline of the model architecture. 
}
\label{fig1}
\end{figure}

To achieve an efficient classification of spiral galaxies, we apply transfer learning (see, e.g., \cite{Pratt1993,Oquab2014}) based on a deep learning model to the color images constructed above. 
Because our previous study (section~3.1 and figure~3 in \cite{Shimakawa2021}) already demonstrated the application of transfer learning to the galaxy images from the HSC-SSP, we briefly summarized the approach herein. 
As is the case with the spiral classification in \citet{Shimakawa2021}, we attached the {\tt Xception} architecture \citep{Chollet2016} of a class of convolutional neural networks (CNN; \cite{Lecun1998,Lecun2015}) to a two-class sigmoid classifier, which is pre-trained for 1000 class separation with the ImageNet dataset \citep{Russakovsky2014}. 
The classifier yields the probability of spiral galaxy ($\hat{y}=[0:1]$) by the 
sigmoid function, through the layers that consists of a dense layer (linear transformation, $d=1024$), batch normalization \citep{Ioffe2015}, and ReLU activation ($\equiv max(0, x)$; \cite{Glorot2011}), as illustrated in figure~\ref{fig1}. 
To implement of this model, we adopted a deep learning model using {\tt TensorFlow} (version 2.6.0; \cite{Abadi2016}) and {\tt Keras} (version 2.6.0; \cite{Chollet2015}), under a single GPU, the NVIDIA TITAN RTX. 

\begin{figure}
\begin{center}
\includegraphics[width=7.5cm]{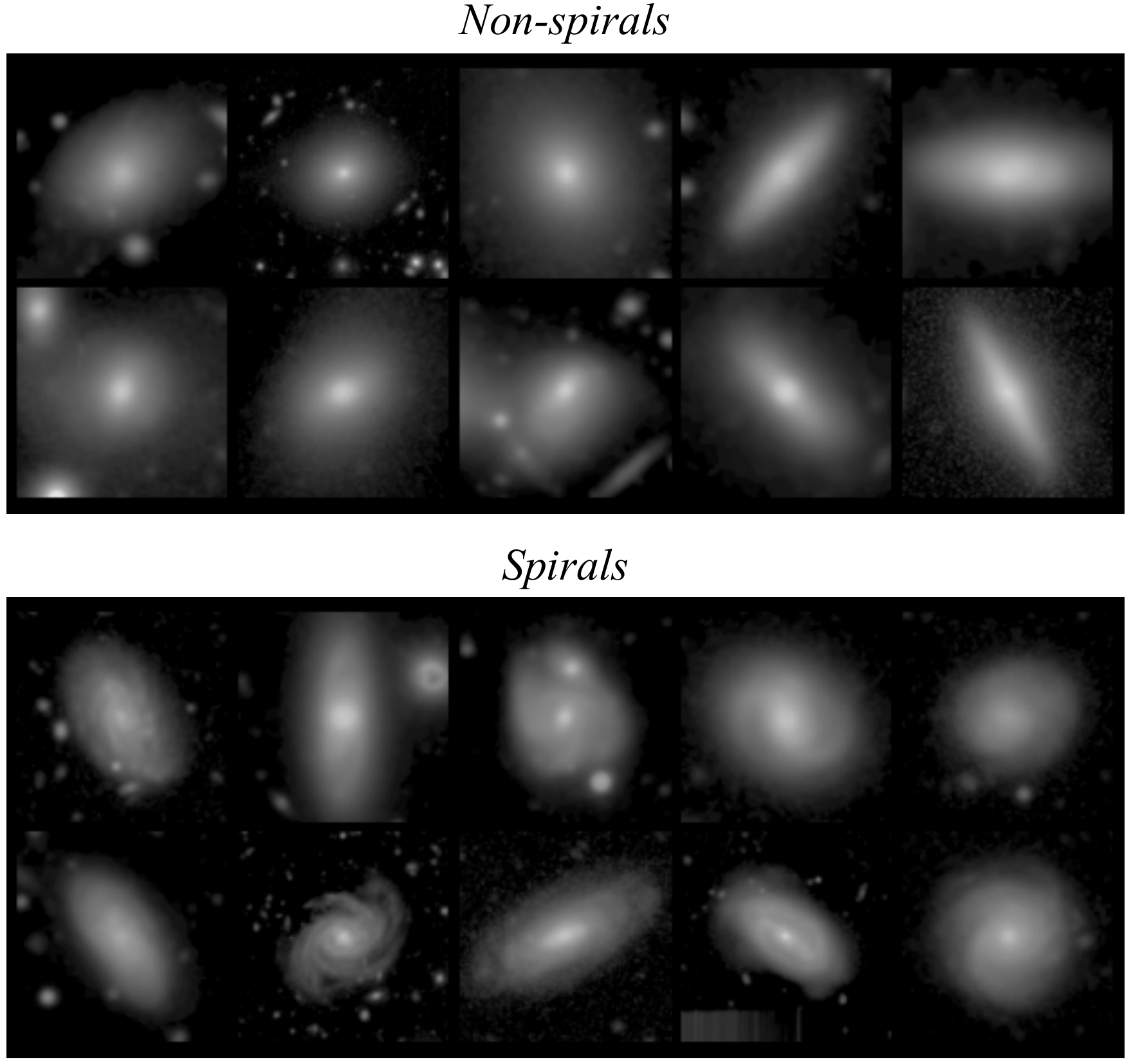}
\end{center}
\caption{
Randomly selected image samples of the training data. 
}
\label{fig2}
\end{figure}

We prepared 6400 training data, 1600 validation data, and 2000 test data for the model training. 
Through the deep learning analysis, we adopted the gray-scale images converted from the RGB color images (figure~\ref{fig2}) to minimize a color--morphology dependence (e.g., \cite{deVaucouleurs1961,Schawinski2014}), although we treated them as the data with three channels to tailor the input dimension of the model architecture.
For each data, half were random non-spiral galaxies, and the other half were spiral galaxies selected by the author's visual inspection (RS). 
Because they were originally based on the HSC-SSP PDR3 sample at $z=$ 0.01--0.3, the datasets partially overlapped with our main targets (7239 out of 54871). 
Their image sizes were scaled and homogenized to $200\times200$ pixels (the typical image size in the sample) from the cutout size of $(4\times\mathrm{R90})^2$ arcsec$^2$ with a pixel width of 0.168 arcsec (figure~\ref{fig1}). 
In addition, the image color channels were standardized similarly as in the pre-training with the ImageNet \citep{Chollet2016} to leverage the pre-trained weight parameters. 

Subsequently, we implemented the model training. 
An optimizer tunes the model weight parameters to minimize the binary cross-entropy loss, which is defined as: 
\begin{equation}
    Loss \equiv -y\log\hat{y} -(1-y)\log(1-\hat{y}),
\label{eq3}
\end{equation}
where $y$ and $\hat{y}$ are the target value $\{0, 1\}$ and the model prediction $[0:1]$, respectively. 
We adopted an adaptive learning rate optimization algorithm ({\tt Adam}; \cite{Kingma2014}) as the optimizer, with a batch size of 32. 
The model training begins with a default learning rate of $lr=1\times10^{-3}$ until the 20th epoch, which was changed to $lr=2\times10^{-4}$ in and after the 20th epoch. 
Further, we served a general data augmentation using {\tt ImageDataGenerator} in the {\tt Keras} library, which performs random image rotation ($<90$ deg) and horizontal and vertical flips. 

\begin{figure}
\begin{center}
\includegraphics[width=7.2cm]{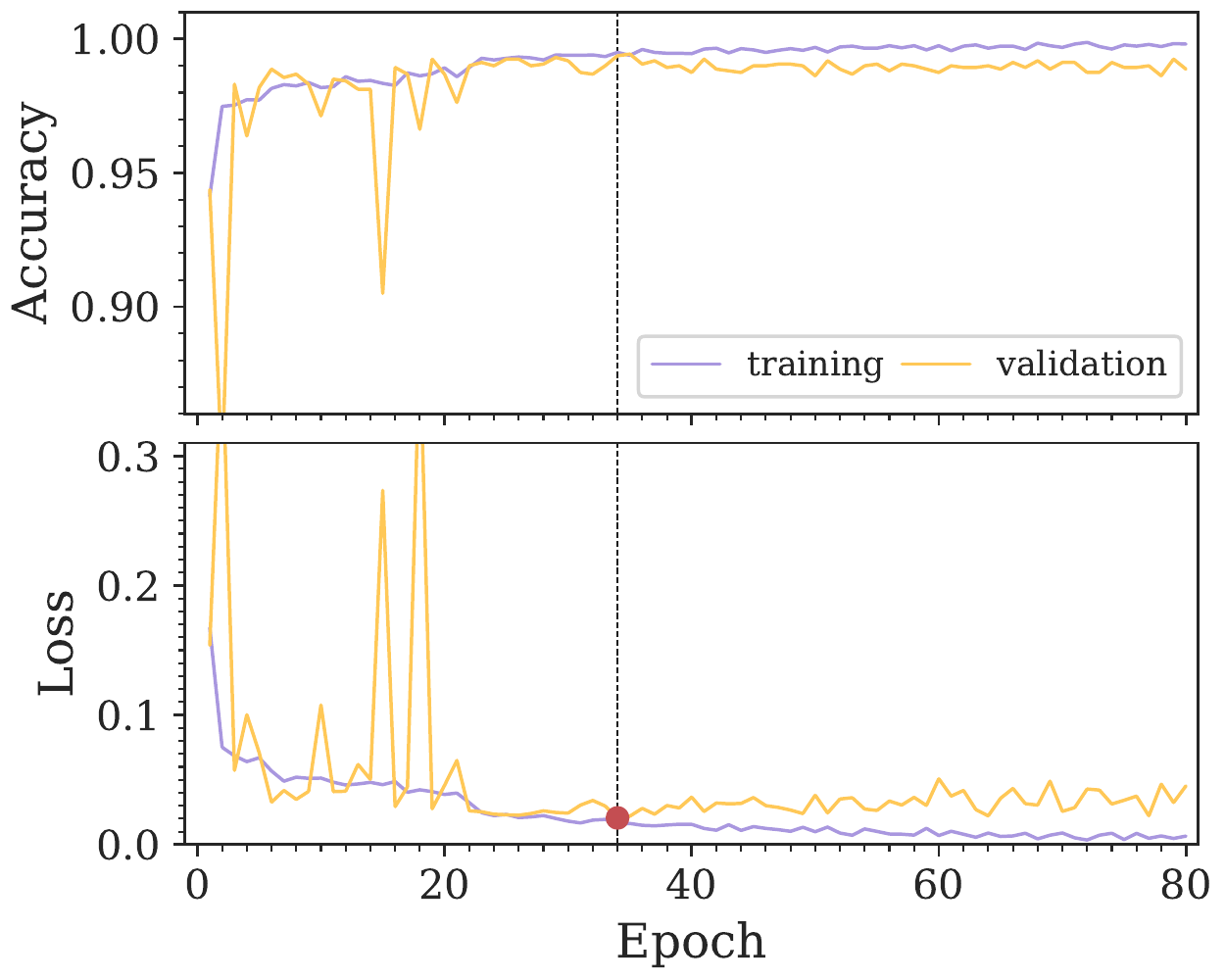}
\end{center}
\caption{
Model training histories in the accuracy (upper panel) and the binary cross-entropy loss (lower panel). 
The purple and yellow lines show the histories of the training and validation data, respectively. 
This study adopts the model in the 34th epoch which has the minimum validation loss as represented by the red-filled circle. 
}
\label{fig3}
\end{figure}

The model learning was conducted in 80 epochs, and it took 5827 sec in our GPU environment. 
The resultant training histories of the model accuracy and binary cross-entropy loss are shown in figure~\ref{fig3}. 
The model accuracy is written as: 
\begin{equation}
    Accuracy \equiv \frac{TP+TN}{TP+TN+FP+FN},
\label{eq2}
\end{equation}
where $TP$ (or $TN$) and $FP$ (or $FN$) indicate the true positive (or negative) and false positive (or negative) rates, respectively. 
Here, the positive and negative classes predicted by the classifier are separated by $\hat{y}=0.5$. 
Among the models in 80 epochs, we chose the best model with the least cross-entropy loss in the validation data (figure~\ref{fig3}). 
The selected model provides a training accuracy of 99.5\%, validation accuracy of 99.4\%, and validation loss of 0.021. 
We also obtained 98.9\% accuracy in independent test data consisting of 1000 spiral and 1000 non-spiral galaxies, which were similarly selected as in the training and validation data.
From the test data, the precision ($TP/(TP+FP)$) and recall ($TP/(TP+FN)$) rates were estimated to be 99.3\% and 98.4\%, respectively.
Moreover, based on the end result (section~\ref{s3}), we noticed the test data include 65 passive spirals.
Although the sample size is small, these passive spirals show the precision rate of 96.9\% and the recall rate of 100\% for the classification of spiral galaxies.

\section{Identifications of passive spiral galaxies}\label{s3}

\subsection{Selection of passive spiral galaxies}\label{s31}

\begin{figure*}
\begin{center}
\includegraphics[width=16.5cm]{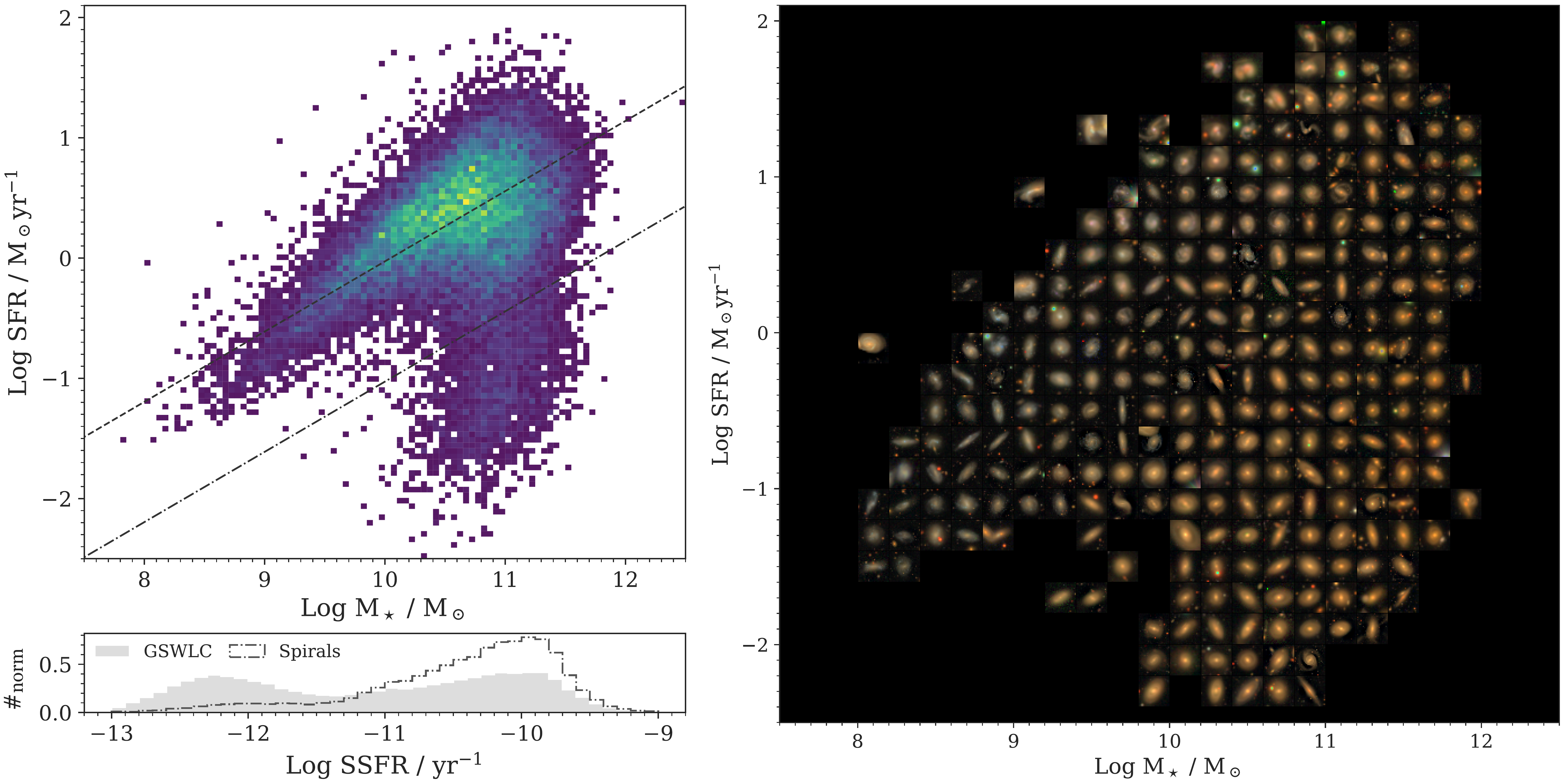}
\end{center}
\caption{
{\it Upper left}: 
Logarithmic SFR versus stellar mass for spiral galaxies selected by the deep learning classification. 
The dashed line shows the best fit star-forming main sequence. 
We selected galaxies lying 1 dex or more below this sequence (dot-dashed line) as candidates of passive spirals. 
{\it Right}: 
Multi-colored $gri$ images of randomly selected spirals on the SFR--M$_\star$ plane. 
{\it Lower left}: 
Distributions of logarithmic specific SFRs of the HSC-SSP PDR3--GSWLC samples (gray histogram) and spiral galaxies (open dot-dashed histogram). 
}
\label{fig4}
\end{figure*}

First of all, we apply the well-tuned deep learning model to the main targets, the 54871 bright galaxies at $z=$ 0.01--0.3, to choose spiral galaxies among them. 
Because our final sample consists of sufficiently bright ($r_\mathrm{petro}<18$) and massive (M$_\star>10^{10}$ M$_\odot$) galaxies at $z<0.3$, we assume that the redshift bias of the spiral classification is negligible (see also \cite{Tadaki2020}). 
Automatic and efficient deep learning classification enabled us to categorize such a large amount of dataset in only 330 sec on the GPU environment. 
Figure~\ref{fig4} represents the frequency distribution and image samples of the selected spirals on the SFR--M$_\star$ plane derived by the GSWLC-2 \citep{Salim2018}. 
This study defines as the spiral galaxies those with model prediction values greater than $\hat{y}=0.5$ (section~\ref{s23}), which amounts to a total of 22720 sources (41\% of the parent sample). 
Because this study solely aims to establish a passive spiral sample, we do not delve into more detailed shape measurements of spiral structures and debiased values of such as the spiral fraction, which are left to future research. 

For the sample classified as spiral galaxies, we choose candidates of passive spirals, located by an order of magnitude lower than the median-fit line for the SFR--M$_\star$ distribution of star-forming galaxies with specific SFRs, SSFR $>10^{-11}$ yr$^{-1}$ (figure~\ref{fig4}):
\begin{equation}
    \log(\mathrm{SFR/M}_\odot\mathrm{yr}^{-1}) < 0.584 \times \log(\mathrm{M}_\star/\mathrm{M}_\odot) - 6.867.
\label{eq4}
\end{equation}
As a result, 2241 (9.9 percent) passive spiral candidates satisfied the SSFR cut, which is significantly lower than a passive fraction of 50 percent in the original sample. 
Within the selected spiral galaxies, we did not find a redshift dependence of the passive fraction.

\subsection{Visual confirmation}\label{s32}

Next, visual inspections for further validation were performed by all authors independently. 
This visual check was quite important for the passive spiral selection because contamination by ringed S0 galaxies to the spiral sample becomes significant in the passive population (see, e.g., \cite{Buta1996,Buta2017}). 
In the validation process, suspicious candidates were tagged by the authors, and then, we calculated a vote fraction for each candidate ({\tt f\_vote = \{0,1\}}; see table~\ref{tab2}).
In conducting the visual classifications, we first established a consensus that obvious pseudo-ringed galaxies \citep{Buta1991,Buta1996,Kormendy2004} were to be regarded as spirals, whereas we tagged visual check flags for non-spirals, merging galaxies, or ambiguous objects that are difficult to determine rings or pseudo-rings. 

\begin{figure*}
\begin{center}
\includegraphics[width=16.5cm]{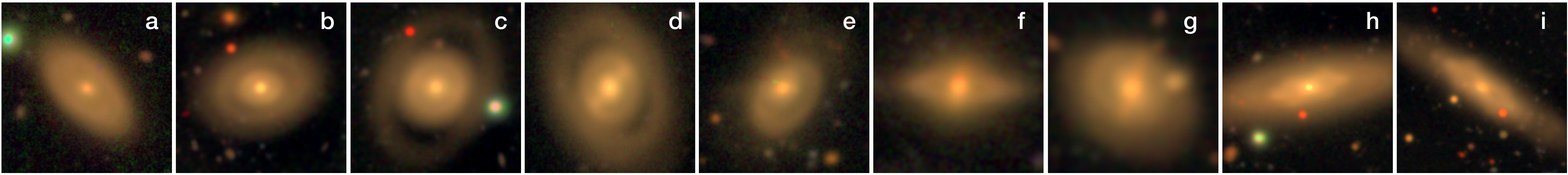}
\end{center}
\caption{
Sample $gri$ images of the candidates judged to have non spiral features by PI's visual inspection.
Most of these suspicious sources are difficult to distinguish as either pseudo-ringed spirals or ringed S0 as seen in (a--d).
For the rest, some sources seem to have unique morphologies, e.g., (e) galaxies with an offset ring such as the Cartwheel Galaxy \citep{Zwicky1941}, (f,g) polar-ring galaxies \citep{Sandage1961}, and (h,i) galaxies with diamond-shaped wings surrounded by an outer ring or disk.
See the text for further explanations on visual checks and human bias.
}
\label{fig5}
\end{figure*}

Figure~\ref{fig5} highlights suspicious samples that were regarded as non-spiral galaxies in the visual validation.
They mostly seem to be ringed or pseudo-ringed galaxies, as shown in figure~\ref{fig5}a-d. 
The remaining suspicious candidates had unclear spiral features or peculiar substructures (figure~\ref{fig5}e-g), for example, an offset ring such as the Cartwheel Galaxy 
\citep{Fosbury1977} and a polar ring \citep{Sandage1961}. 
Additionally, we found two unique objects with diamond-shaped wings ({\tt id\_hsc3 = 41033924272474055 \& 41034615762212336}; table~\ref{tab2} and figure~\ref{fig5}h,i), which were also removed from the passive spiral sample. 
They have a similar structure to local galaxies, NGC~7020 \citep{Buta1990}, NGC~4429 \citep{Buta2007}, and IC~4767 \citep{Whitmore1988}, which are reported as peculiar S0 galaxies with hexagonal or X shapes. 
It would be intriguing to investigate their kinematics and accretion histories using a modern integral field spectrograph, which will address the formation mechanisms of such a peculiar substructure.

Table~\ref{tab1} summarizes the results of our visual inspections.
To quantify the human bias in our visual validations, we define a concordance rate (CR) by:
\begin{equation}
    \mathrm{CR} \equiv N_s(A|B) = N_s(A\cap B)/N_s(B),
\label{eq5}
\end{equation}
where $N_s$($x$) is the number of suspicious objects (i.e., non passive spirals) among the candidates ($N=2241$) judged by an author $x$; therefore, CR indicates that the number ratio of overlapped non passive spirals in between authors A and B ($=N_s$(A$\cap$B)) to that in the author B ($=N_s$(B)).
The derived CR values between individual authors are also summarized in table~\ref{tab1}.
The table indicates the significant variations in our visual check results ($N_s=$ 570--1466) and concordance rates, i.e., human bias, even among experts.
For example, the author 5 selected the least ($N_s=570$) as non spiral features from the candidates, most of which ($\sim90$ percent) were selected by the author 1, 2 and 6 as well.
On the other hand, 24--31 percents of non passive spirals deemed by the author 5 were passed in visual validations by the author 3, 4 and 7.
Meanwhile, the authors 1, 2 and 6 who selected more samples ($N>1300$) as non-spiral galaxies than the others, with $\sim80$ percent in agreement.
Such significant variations in our visual validations reflect difficulties in the visual classification of passive spiral galaxies.
These inconsistencies would be mostly due to the difficulty of distinguishing pseudo-ringed spiral galaxies from ringed S0 galaxies or interacting objects with tidal tales, even with high-quality data from the Subaru HSC.
The viewing angles of galaxies in the sky would be another factor causing the variations.

\begin{figure}
\begin{center}
\includegraphics[width=7.5cm]{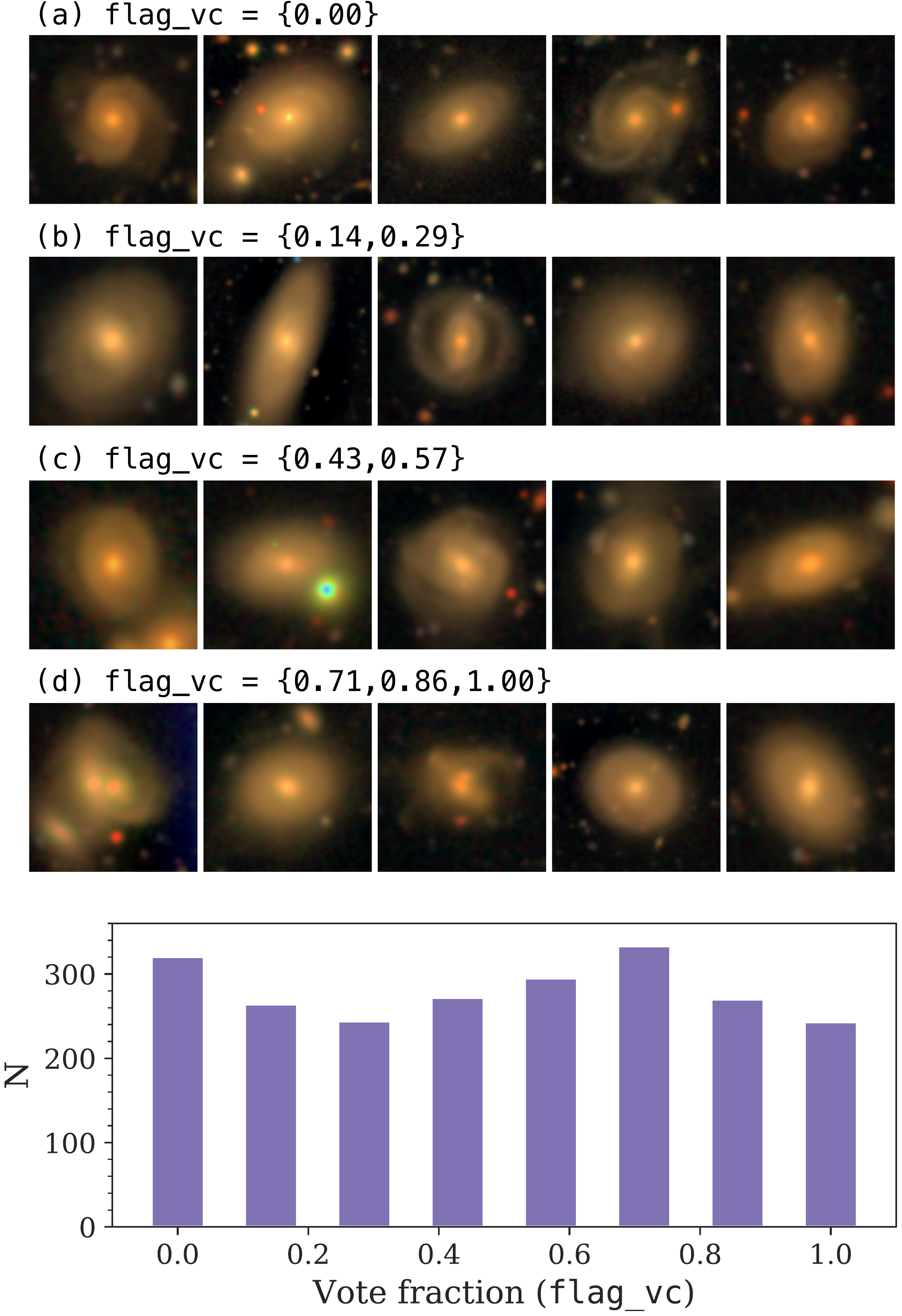}
\end{center}
\caption{
The bottom histogram shows the vote fraction ({\tt flag\_vc}) for suspicious sources in the passive spiral candidates selected by the deep learning model. 
Color postage stamps are random examples given vote fractions: from the top, (a) {\tt flag\_vc = 0}, that is, they are recognized as spirals by every co-authors, (b) {\tt flag\_vc = 0.14 or 0.29}, (c) {\tt flag\_vc = 0.43 or 0.57}, (d) {\tt flag\_vc = 0.71, 0.86, or 1}.  
}
\label{fig6}
\end{figure}

Based on the responses of our visual inspections, we calculated the vote fraction for each object ({\tt flag\_vc = [0:1]}; table~\ref{tab2}), which provides flexibility for catalog users to make their decisions on the sample selection from the passive spiral catalog (table~\ref{tab2}).
Higher {\tt flag\_vc} values indicate less reliable passive spiral samples in our visual validations.
Image samples for each vote fractions are represented in figure~\ref{fig6}, showing that the samples with higher {\tt flag\_vc} values tend to have less clear spiral features or ring-like morphologies.
If one wants to establish the most secure passive spiral sample from our catalog, we recommend using the sources with {\tt flag\_vc = 0}, which are recognized as having spiral features by all authors of this paper.
However, it should be noted that such a conservative selection may overlook many pseudo-ringed spiral galaxies.

\begin{table*}
\caption{
Concordance rates ($\equiv N_s$(A$|$B)) among visual inspections by 7 authors (AU1--7).
The 2nd row indicates the number of the candidates that are judged to have no spiral features by each co-author ($N_s$).
}
\label{tab1}
\centering
\begin{tabular}{r|rrrrrrr}
    \backslashbox{A}{B} & AU1 & AU2 & AU3 & AU4 & AU5 & AU6 & AU7\\ \hline
    N$_s$ & 1357 & 1344 & 860 & 1066 & 570 & 1466 & 1071\\\hline
    AU1 & ... & 0.81 & 0.92 & 0.77 & 0.92 & 0.77 & 0.83\\
    AU2 & 0.81 & ... & 0.87 & 0.77 & 0.90 & 0.75 & 0.77\\
    AU3 & 0.58 & 0.56 & ... & 0.53 & 0.76 & 0.50 & 0.55\\
    AU4 & 0.61 & 0.61 & 0.66 & ... & 0.74 & 0.56 & 0.65\\
    AU5 & 0.39 & 0.38 & 0.50 & 0.39 & ... & 0.35 & 0.37\\
    AU6 & 0.83 & 0.82 & 0.85 & 0.77 & 0.91 & ... & 0.84\\
    AU7 & 0.65 & 0.61 & 0.69 & 0.65 & 0.69 & 0.61 & ...\\
\end{tabular}
\end{table*}

\subsection{Passive spiral galaxy catalog}\label{s33}

\begin{figure*}
\begin{center}
\includegraphics[width=15cm]{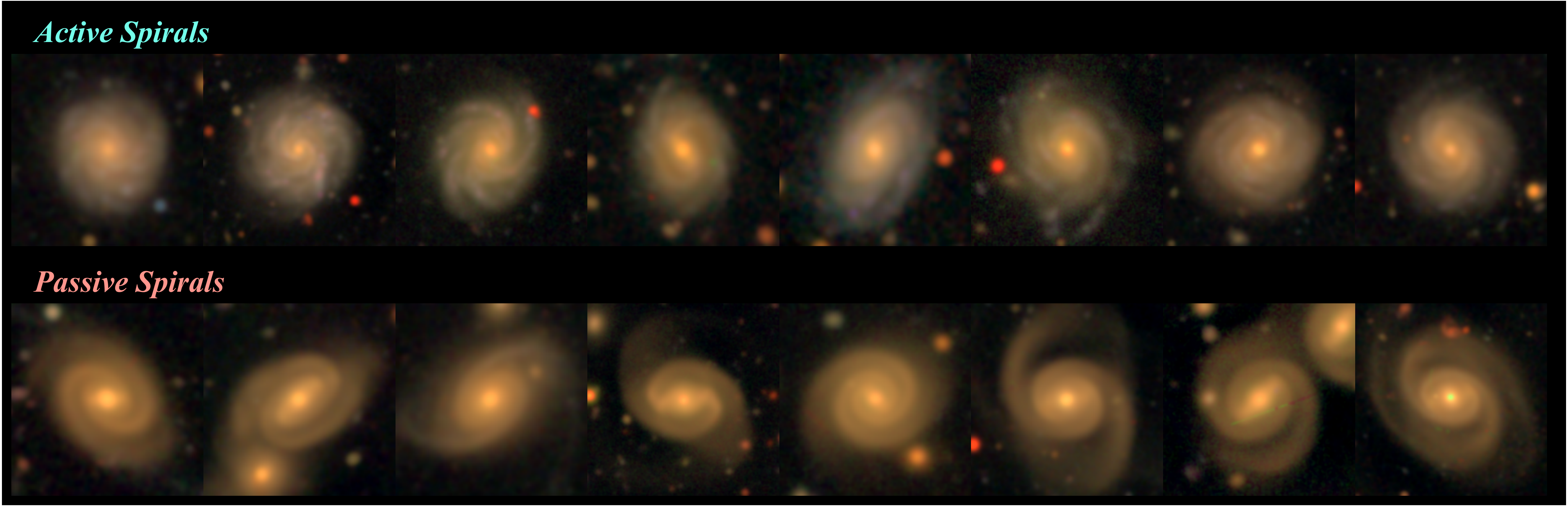}
\end{center}
\caption{
Example RGB color images of active (top) and passive (bottom) spiral galaxies at $z=$ 0.01--0.3 based on $gri$ cutouts ($4~\mathrm{R90}\times4~\mathrm{R90}$ arcsec$^2$) from the HSC-SSP PDR3. 
Full postage stamps of our passive spiral samples are available as supplemental material. 
}
\label{fig7}
\end{figure*}

Our passive spiral catalog (table~\ref{tab2}) and individual postage stamps in {\tt png} format can be downloaded on the HSC-SSP PDR3 website (\url{https://hsc.mtk.nao.ac.jp/ssp/data-release/}).
Throughout of this paper, we employ 1100 candidates with {\tt flag\_vc} $<0.5$ as the passive spiral galaxies, that is, we adopt the samples accepted by 4 or more out of 7 authors of this paper in the last section~\ref{s32}.
The passive fraction of spiral galaxies after the visual validations is 5.1\% (from $1100/(22720-(2241-1100))$)\footnote{We here ignore the contaminants in the active spiral samples, given the high success rate of the overall spiral classification (section~\ref{s23}).}, which is the same level as the red fraction of spirals (4--8\%) in the {\tt Galaxy Zoo} \citep{Masters2010}.
Whereas, we note that this consistency includes both reducing and increasing effects from more conservative passive selection using UV-to-IR data \citep{Salim2018} and higher quality HSC imaging data \citep{Aihara2021}, respectively.
Example $gri$-composite color images of star-forming (active) and passive spiral galaxies are shown in figure~\ref{fig7}, which clearly differ in appearance: passive spirals are redder with smoother spiral arms than active spirals.
All the color images of the selected passive spirals are summarized by supplemental material available online.
Postage stamps of the secondary samples with {\tt flag\_vc} values ($>0.5$), which are not discussed hereafter, are also available as supplemental material.

\begin{figure*}
\begin{center}
\includegraphics[width=15cm]{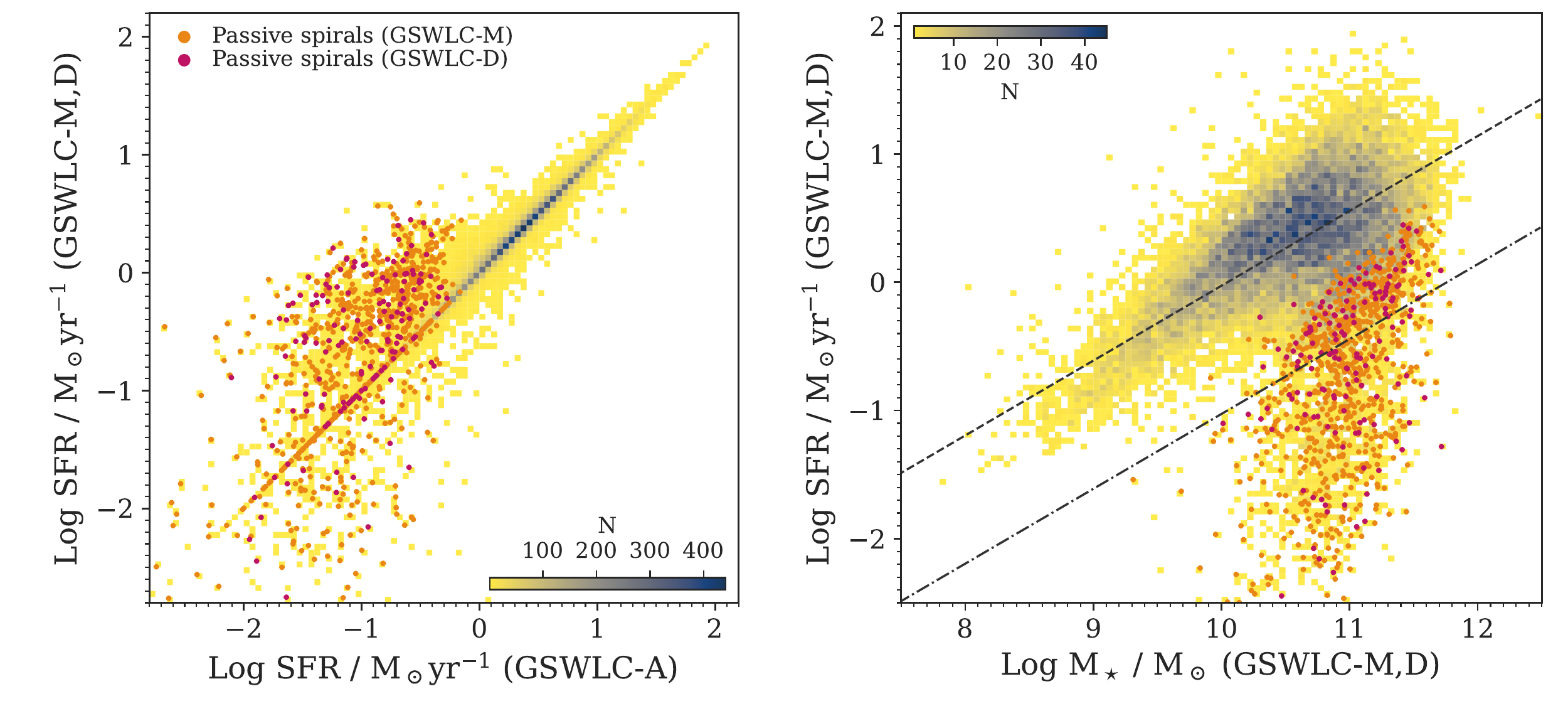}
\end{center}
\caption{
The left panels shows a comparison of logarithmic SFR based on GSWLC-A with those from GSWLC-M or -D \citep{Salim2018}, and the right panel represents revised logarithmic SFR--M$_\star$ based on GSWLC-M or -D from figure~\ref{fig4}.
The color map shows the distribution of all spiral galaxies available in GSWLC-M or -D. 
The orange and red circles indicate the passive spiral samples that are available in GSWLC-M or -D (but selected with GSWLC-A), respectively. 
While they are mostly equal to each other, there are significant outliers. 
}
\label{fig8}
\end{figure*}

Moreover, we checked if they remained quiescent even when we adopted the best-fit physical properties based on deeper GALEX data if available in the version of the GSWLC-M or GSWLC-D (see \cite{Salim2016,Salim2018}).
It turns out that 786 (818) and 158 (154) objects out of 1100 main (1141 secondary) samples were covered by GSWLC-M or GSWLC-D, respectively.
As noted in the literature, the depth of the UV data influences the SFR measurement, especially for less-active galaxies. 
Once we replaced their physical values with deeper UV data, many of the passive spiral samples were located in the green valley on the star-forming main sequence (figure~\ref{fig8}) and approximately half of them (339 and 82 out of 786 and 158 in GSWLC-M and -D, respectively) were shifted into above the threshold of the passive population (Eq.~\ref{eq4}).
This indicates that only deep optical data are inadequate to select truly passive spiral galaxies and sufficiently deep UV (and IR) data are crucial (see also \cite{Fraser-McKelvie2016}).
We note that no significant difference existed in the behavior of between SFRs from GSWLC-M and -D to those from GSWLC-A.
Consequently, our catalog provides an additional validation flag ({\tt flag\_uv}), which yields ``M" or ``D" if the candidates still satisfy the selection threshold even if we adopt the catalog versions of GSWLC-M or -D, respectively.
However, we should note that this paper does not adopt {\tt flag\_uv} to maintain the homogeneous selection of passive spirals over the entire survey field.

Combined with the visual check flag ({\tt flag\_vc}), we offer a comprehensive and flexible source catalog (table~\ref{tab2}) to leave further decisions up to users.
These validation flags are aimed at allowing users of our catalog to select more (or less) secure passive spiral sources for follow-up investigations, depending on their objectives.

\begin{table}
\caption{
Column descriptions of the passive spiral catalog. 
The full catalog is available on the HSC-SSP PDR3 site (\url{https://hsc-release.mtk.nao.ac.jp}). 
}
\label{tab2}
\begin{flushleft}
	\begin{tabular}{ll} 
	\hline
	Column         & Description\\
	\hline
	{\tt id\_hsc3} & object\_id in HSC-SSP PDR3$^1$\\
	{\tt id\_sdss} & object\_id in SDSS (GSWLC)$^2$\\
	{\tt ra}       & right ascension (deg)\\
	{\tt dec}      & declination (deg)\\
	{\tt redshift} & spectroscopic redshift from SDSS DR15$^3$\\
	{\tt y\_pred}  & probability value by CNN classification\\
	{\tt flag\_vc} & vote fraction for suspicious objects$^4$\\
	{\tt flag\_uv} & validation flag with in-depth UV data$^5$\\
	\hline
	\end{tabular}
\end{flushleft}
\footnotesize{$^1$ Refer to \citet{Aihara2021}}\\
\footnotesize{$^2$ Cross-match within 1 arcsec to \citet{Salim2018}}\\
\footnotesize{$^3$ Cross-match to \citet{Aguado2019} on the HSC-SSP database}\\
\footnotesize{$^4$ See section~\ref{s32} for details.}\\
\footnotesize{$^5$ `M' or `D' if sources remain passive even in the 
SFR--M$_\star$ plain based on GSWLC-M or GSWLC-D (see section~\ref{s31}).}
\end{table}

\section{Discussion}\label{s4}

Based on 54871 spec-$z$ sources at $z=0.01{\rm -}0.3$ from the HSC-SSP PDR3 and the GSWLC-2, we established a passive spiral sample, in which 1100 objects are handled here as the main sample. 
This section discusses the physical properties and environmental effects using EW$_\mathrm{H\delta}$--D$_n$4000 and phase-space diagrams, respectively.  

\subsection{EW$_\mathrm{H\delta_A}$--D$_n$4000 diagram}\label{s41}

Investigation of the stellar populations of passive spirals enables us to understand their evolutionary stages compared to those of the other galaxies. 
We examined the 4000~\AA\ break strength (D$_n$4000) and the Balmer absorption line index (H$\delta$ equivalent width, EW$_\mathrm{H\delta_A}$; \cite{Worthey1997}) of passive spirals, and compared them with those of spiral galaxies and passive galaxies. 
The so-called EW$_\mathrm{H\delta_A}$--D$_n$4000 diagram has been widely used to understand stellar populations of quiescent and post-starburst galaxies \citep{Blake2004,Yesuf2014,Wu2021}. 
While the D$_n$4000 is roughly interpreted as the ages of stellar populations, the EW$_\mathrm{H\delta}$ responds to the time scale of star formation history, which is the strongest ($>5$ \AA) in 0.1--1 Gyr after starburst \citep{Kauffmann2003,Wu2021}. 

\begin{figure}
\begin{center}
\includegraphics[width=7.5cm]{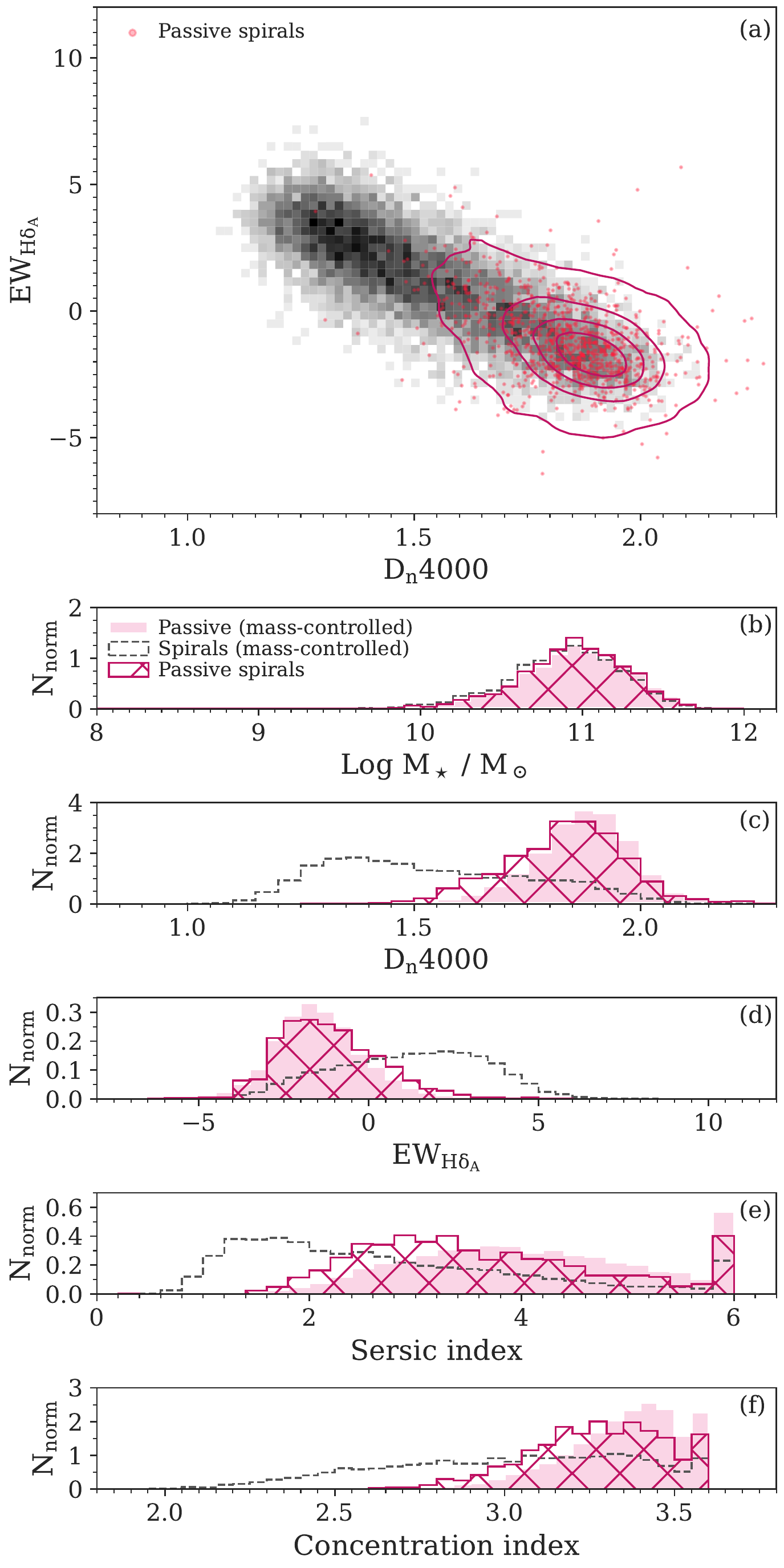}
\end{center}
\caption{
(a) The top panel shows the EW$_\mathrm{H\delta_A}$--D$_n$4000 diagram for our samples with counterparts in the MPE-JHU catalog within a radius of 1 arcsec \citep{Kauffmann2003}. 
The black region and the red contours show distributions of all cross-matched spirals and passive spirals, respectively. 
(b) The second row indicates stellar mass distributions of randomly selected mass-controlled passive galaxies (pink-filled) and spirals (dashed), $N=$ 10000 for each, from the parent sample, and passive spirals (red hatched). 
Panels (c--f) represent normalized distributions of the D$_n$4000, EW$_\mathrm{H\delta_A}$, Sersic index, and Petrosian concentration index, respectively. 
The concentration measurements are based on the NYU-VAGC catalog \citep{Blanton2003,Blanton2005}. 
The sample in each histogram corresponds to those in the panel (b). 
}
\label{fig9}
\end{figure}

We obtained D$_n$4000 and EW$_\mathrm{H\delta_A}$ values of our sample from the MPE-JHU catalog \citep{Kauffmann2003} by searching for their counterparts within a 1 arcsec radius. 
Among the 54871 samples, 50831 objects, including 970 passive spirals, are cross-matched with the MPE-JHU catalog. 
The density distribution on the D$_n$4000 and EW$_\mathrm{H\delta_A}$ diagram and normalized distributions in each axis for spiral galaxies, passive galaxies, and passive spirals are shown in figure~\ref{fig9}a.

As a result, we found that passive spiral galaxies have a distribution similar to that of general passive galaxies more than active spirals on the D$_n$4000 and EW$_\mathrm{H\delta_A}$ plane (figure~\ref{fig9}c,d), where we randomly selected 10000 mass-controlled passive galaxies and active spirals from the parent samples to adjust the stellar mass distributions to that of passive spirals (figure~\ref{fig9}b). 
We confirmed that such a trend becomes more evident when we employed more secure passive spirals from deeper UV data ({\tt flag\_uv = M or D}; table~\ref{tab2}). 
Because we selected passive galaxies with the same threshold on the SFR--M$_\odot$ plane (figure~\ref{fig4}), the similar distributions between these two samples suggest that the passive spirals have identical characteristics to the typical quiescent galaxies in stellar populations despite the significant difference in their morphologies. 
Distributions of Sersic \citep{Sersic1968} and Petrosian concentration indices \citep{Shimasaku2001} for passive spiral galaxies in the SDSS $r$-band are more similar to those of typical quiescent galaxies than active spirals. 
These concentration measurements were taken from the NYU-VAGC catalog \citep{Blanton2003,Blanton2005} by spatial cross-matching within 1 arcsec (52166 sources, including 1009 passive spirals, were successfully matched). 

The results may suggest that star formation histories and galaxy morphology transitions are not always coupled, and the quenching timescales of spiral galaxies are very slow ($\sim$ a few Gyr), as proposed by \citet{Schawinski2014}. 
However, it should be noted that the SDSS spectra mostly trace bright central bulges of galaxies within the fibers of 3 arcsec diameter, which is smaller than the typical Petrosian half-light radius of 1.76 arcsec in our sample.
The concentration metrics show that passive spirals tend to have bulge structures similar to those of typical quiescent (non-spiral) galaxies; thus, the similar stellar populations between these spectra are not surprising.
An extensive deep survey with an integral field spectrograph is needed to reach a consensus on the star formation histories of passive spiral arms.

\subsection{Phase-space diagram}\label{s42}

Previous studies have reported a higher frequency of passive spirals at higher local densities \citep{vandenBergh1976,Goto2003,Masters2010}, suggesting that environmental effects may be responsible for such a high passive fraction of spiral galaxies. 
This study discusses the environmental quenching of spiral galaxies from a more practical standpoint by leveraging a large passive spiral sample over 1072 square degrees. 
Specifically, we cross-matched the main targets at $z=$0.01--0.3 (section~\ref{s21}) with X-ray-selected clusters ($\sim10^{14}{\rm -}10^{15}$ M$_\odot$) by SPIDERS (the SPectroscopic IDentification of eROSITA Sources; \cite{Clerc2016}) to examine the dynamical associations of passive spirals with massive clusters. 
Because the SPIDERS catalog offers virial radii ($R_{200}$) and velocity dispersion ($\sigma_v$), we can easily test the association of passive spirals with clusters on the phase-space diagram \citep{Bertschinger1985} by stacking all individually matched cluster members to the cluster sources. 
The phase-space diagram provides a useful clue on the dynamical states of galaxy clusters and the trajectory of galaxies infalling towards the cluster center \citep{Sanchis2004,Mahajan2011,Jaffe2015,Oman2016,Yoon2017}. 
In this study, the phase-space diagram enables us to infer where spiral galaxies are inclined to be passive during their infall into the clusters. 

\begin{figure}
\begin{center}
\includegraphics[width=7.5cm]{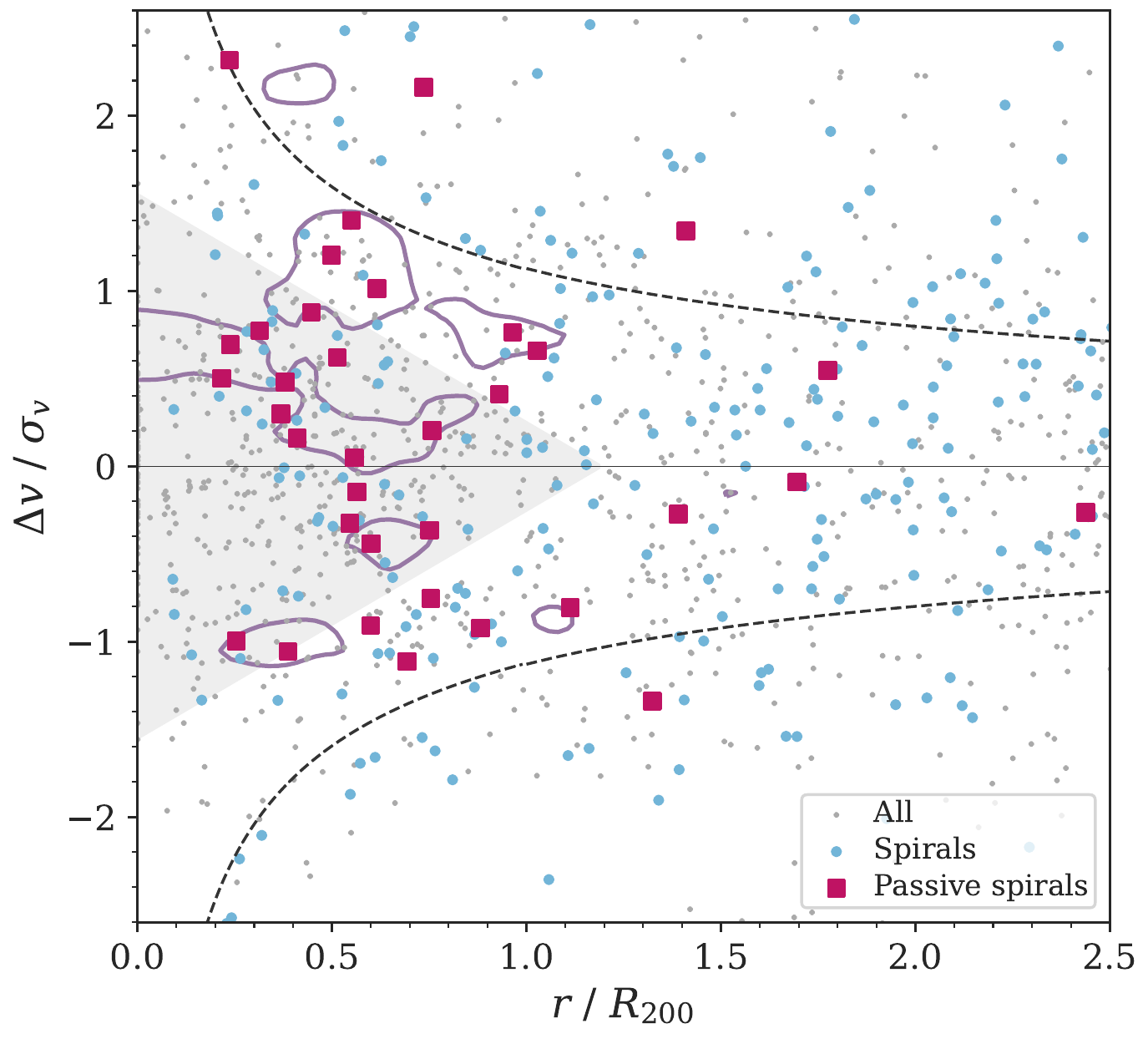}
\end{center}
\caption{
The phase-space diagram for our samples (gray dots) associated with X-ray clusters ($>10^{14}$ M$_\odot$; \cite{Clerc2016}). 
The blue circles and red squares are spirals and passive spiral galaxies, respectively.
The purple contours indicate peak areas of the passive fraction of spiral galaxies ($\ge40$\%) on the diagram, reaching up to $\sim50$\%.
The dashed curves and gray regions are the escape velocity and virialized area, respectively \citep{Jaffe2015}.
}
\label{fig10}
\end{figure}

We obtained 1148 cluster members associated with 123 clusters within the scaled projected radii ($r/R_{200}<2.5$) and velocity offsets ($|\Delta v/\sigma_{v}|<2.6$).
Among the matched cluster members, 250 and 35 sources were classified as spiral and passive spiral galaxies, respectively. 
The average passive fraction of spiral galaxies in the clusters (14\%) was approximately thrice as high as that in random fields (5\%).
Figure~\ref{fig10} shows the positions of all cluster members on the projected phase-space diagram. 
The figure also indicates the escape velocity and virialized region proposed by \citet{Jaffe2015}, defining the orbital states on the projected phase-space diagram more realistically than in previous studies. 

The very center of the galaxy clusters lacks spiral galaxies (figure~\ref{fig10}), which is well known as the morphology--density relation (e.g., \cite{Dressler1980,Cappellari2011}); we also obtained a non-uniform distribution of passive spirals to the entire spiral galaxies in the cluster environments. 
We detected a substantial excess of passive spiral galaxies reaching up to $\sim50$\% on this diagram in the area of $r/R_{200}=0.3{\rm -}1$ and $|\Delta v/\sigma_{v}|\lesssim1$, where galaxies are expected to experience core crossing more than once (e.g., \cite{Jaffe2015,Yoon2017}). 
It is quite reasonable to see such a biased distribution owing to the truncation of the continuous H{\sc i} gas supply due to the ram pressure stripping in the earlier infall phase \citep{Hamabata2019}. 
Previous studies have reported that spiral galaxies influenced by the ram pressure stripping were found in the 1st infall ($|\Delta v/\sigma_{v}|=1{\rm -}2.5$ and $r/R_{200}<1.3$) and backsplash ($|\Delta v/\sigma_{v}|\le0.5$ and $r/R_{200}=1.2{\rm -}2$) areas (see, e.g., \cite[figure~5]{Yoon2017}). 
Thus, our results provide strong independent support for the conclusions presented in previous studies that ram pressure stripping enhances the fraction of passive spirals in cluster environments \citep{Bothun1980,Giovanelli1983,Cayatte1994,Bravo-Alfaro2001,Elmegreen2002,Boselli2006}. 
We also observed some passive spirals in cluster outskirts ($r>R_{200}$), which may have been formed through the pre-processing in falling subclusters \citep{Fujita2004,Rhee2020}. 

Finally, the environmental effects would not be the only causal factors that transform star-forming spirals into passive spirals because passive spirals are observed in low-density environments and in the vicinity of galaxy clusters and groups (see also \cite{Masters2010}). 
Nevertheless, our results suggest that the environmental effects can extend the passive spiral phase and/or help spiral galaxies to be quiescent. 

\section{Conclusion}\label{s5}

Based on the deep learning morphology classification for the HSC-SSP PDR3 imaging data, we constructed 1100 passive spiral samples at $z<0.3$ over 1072 deg$^2$, which are in a quiescent phase as inferred from their UV to mid-IR SEDs by the GSWLC-2 catalog \citep{Salim2016,Salim2018}. 
As a result, we obtained 1100 relatively reliable passive spirals and 1141 secondary samples consisting of ringed S0, pseudo-ringed galaxies, and false detection or peculiar types through visual inspection. 
Validation flags based on visual inspections by authors and selections with deeper UV data will allow users to flexibly select more (or less) secure passive spirals.
All data related to this study are publicly available on the data release site of the HSC-SSP PDR3 (\url{https://hsc.mtk.nao.ac.jp/ssp/data-release/}).
Our large passive galaxy sample will help ones to delve into more detailed properties of passive spirals in individuals to unveil the late evolutionary phase of spiral galaxies. 
Some secondary samples have unique structures, such as polar rings and diamond wings. 
While these special morphological types are beyond the scope of this study, they would provide us with further insight into the formation mechanisms of galaxy morphologies. 

We also found that the characteristics of stellar populations in passive spirals inferred from the EW$_\mathrm{H\delta_A}$ and D$_n$4000 indices were similar to those of typical quiescent galaxies. 
Further research is required to understand the basis for such morphological discrepancies between general quiescent galaxies and passive spirals; 
for instance, IFU observations in the optical regime \citep{Ishigaki2007} and high-resolution cold gas observation with the Atacama Large Millimeter/submillimeter Array (ALMA) will be helpful in resolving the detailed kinematics and consumption processes of cold gas. 
In addition, we investigated the kinematic positions of passive spirals associated with X-ray clusters on the phase-space diagram. 
We then detected the peak fraction of the passive spirals over spiral galaxies around the midterm to late infall phase on the phase-space plane, supporting the ram pressure scenario, which has been widely suggested in previous studies \citep{Bothun1980,Giovanelli1983,Cayatte1994,Bravo-Alfaro2001,Elmegreen2002,Boselli2006}. 

The current passive spiral sample still relies highly on visual inspection, especially to divide passive spirals with ringed S0 galaxies. 
Such a time-consuming procedure should be automated when the large next-gen datasets are delivered by such as the LSST project on the Vera C. Rubin Observatory \citep{Ivezic2019} and the High Latitude Wide Area Survey with the Nancy Grace Roman Space Telescope \citep{Spergel2015}. 
To achieve a more effective classification, the citizen science project dedicated to the HSC data, the {\tt GALAXY CRUISE} program, will be quite useful for the creation of the training data, which provides us with not only the spiral classification but also the presence or absence of the ring feature. 
Thus, future updates combined with the {\tt GALAXY CRUISE} data will enable us to conduct more effective and comprehensive classifications.


\begin{ack}
Based on data collected at the Subaru Telescope and retrieved from the HSC data archive system, which is operated by Subaru Telescope and Astronomy Data Center at National Astronomical Observatory of Japan. We are honored and grateful for the opportunity of observing the Universe from Maunakea, which has the cultural, historical and natural significance in Hawaii. 

The Hyper Suprime-Cam (HSC) collaboration includes the astronomical communities of Japan and Taiwan, and Princeton University. The HSC instrumentation and software were developed by the National Astronomical Observatory of Japan (NAOJ), the Kavli Institute for the Physics and Mathematics of the Universe (Kavli IPMU), the University of Tokyo, the High Energy Accelerator Research Organization (KEK), the Academia Sinica Institute for Astronomy and Astrophysics in Taiwan (ASIAA), and Princeton University. Funding was contributed by the FIRST program from Japanese Cabinet Office, the Ministry of Education, Culture, Sports, Science and Technology (MEXT), the Japan Society for the Promotion of Science (JSPS), Japan Science and Technology Agency (JST), the Toray Science Foundation, NAOJ, Kavli IPMU, KEK, ASIAA, and Princeton University. 
This paper makes use of software developed for the Large Synoptic Survey Telescope. We thank the LSST Project for making their code available as free software at  \url{http://dm.lsst.org}.

The Pan-STARRS1 Surveys (PS1) have been made possible through contributions of the Institute for Astronomy, the University of Hawaii, the Pan-STARRS Project Office, the Max-Planck Society and its participating institutes, the Max Planck Institute for Astronomy, Heidelberg and the Max Planck Institute for Extraterrestrial Physics, Garching, The Johns Hopkins University, Durham University, the University of Edinburgh, Queen’s University Belfast, the Harvard-Smithsonian Center for Astrophysics, the Las Cumbres Observatory Global Telescope Network Incorporated, the National Central University of Taiwan, the Space Telescope Science Institute, the National Aeronautics and Space Administration under Grant No. NNX08AR22G issued through the Planetary Science Division of the NASA Science Mission Directorate, the National Science Foundation under Grant No. AST-1238877, the University of Maryland, and Eotvos Lorand University (ELTE) and the Los Alamos National Laboratory.

Funding for the Sloan Digital Sky Survey IV has been provided by the Alfred P. Sloan Foundation, the U.S. Department of Energy Office of Science, and the Participating Institutions. SDSS-IV acknowledges support and resources from the Center for High Performance Computing  at the University of Utah. The SDSS website is \url{www.sdss.org}.

SDSS-IV is managed by the Astrophysical Research Consortium for the Participating Institutions of the SDSS Collaboration including the Brazilian Participation Group, the Carnegie Institution for Science, Carnegie Mellon University, Center for Astrophysics | Harvard \& Smithsonian, the Chilean Participation Group, the French Participation Group, Instituto de Astrof\'isica de Canarias, The Johns Hopkins University, Kavli Institute for the Physics and Mathematics of the Universe (IPMU) / University of Tokyo, the Korean Participation Group, Lawrence Berkeley National Laboratory, Leibniz Institut f\"ur Astrophysik Potsdam (AIP),  Max-Planck-Institut f\"ur Astronomie (MPIA Heidelberg), Max-Planck-Institut f\"ur Astrophysik (MPA Garching), Max-Planck-Institut f\"ur Extraterrestrische Physik (MPE), National Astronomical Observatories of China, New Mexico State University, New York University, University of Notre Dame, Observat\'ario Nacional / MCTI, The Ohio State University, Pennsylvania State University, Shanghai Astronomical Observatory, United Kingdom Participation Group, Universidad Nacional Aut\'onoma de M\'exico, University of Arizona, University of Colorado Boulder, University of Oxford, University of Portsmouth, University of Utah, University of Virginia, University of Washington, University of Wisconsin, Vanderbilt University, and Yale University.

We thank anonymous referee for helpful feedback.
CB gratefully acknowledges the support of a post-doctoral scholarship from the Natural Sciences and Engineering Research Council of Canada.
PFW acknowledges the support of the fellowship from the East Asian Core Observatories Association. 
We would like to thank Editage (\url{www.editage.com}) for English language editing.

This work made extensive use of the following tools, {\tt NumPy} 
\citep{Harris2020}, {\tt Matplotlib} \citep{Hunter2007}, the Tool for OPerations 
on Catalogues And Tables, {\tt TOPCAT} \citep{Taylor2005}, a community-developed 
core Python package for Astronomy, {\tt Astopy} \citep{AstropyCollaboration2013}, 
and Python Data Analysis Library {\tt pandas} \citep{Mckinney2010}. 
\end{ack}



\bibliographystyle{pasj}
\bibliography{rs21c} 

\begin{figure*}
\centering
\includegraphics[width=17cm]{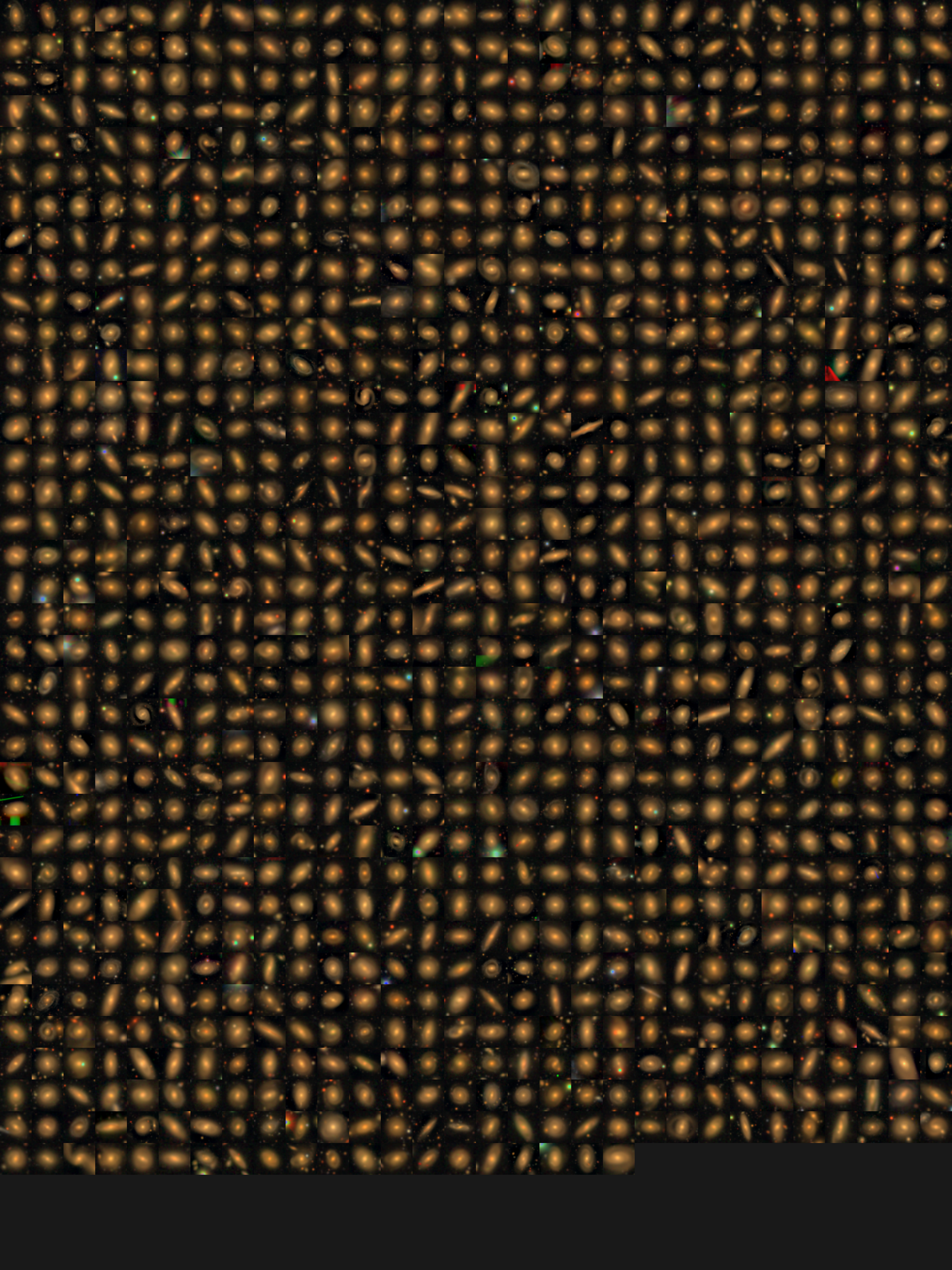}
\caption{
The main samples.
}
\label{fig1a}
\end{figure*}

\begin{figure*}
\centering
\includegraphics[width=17cm]{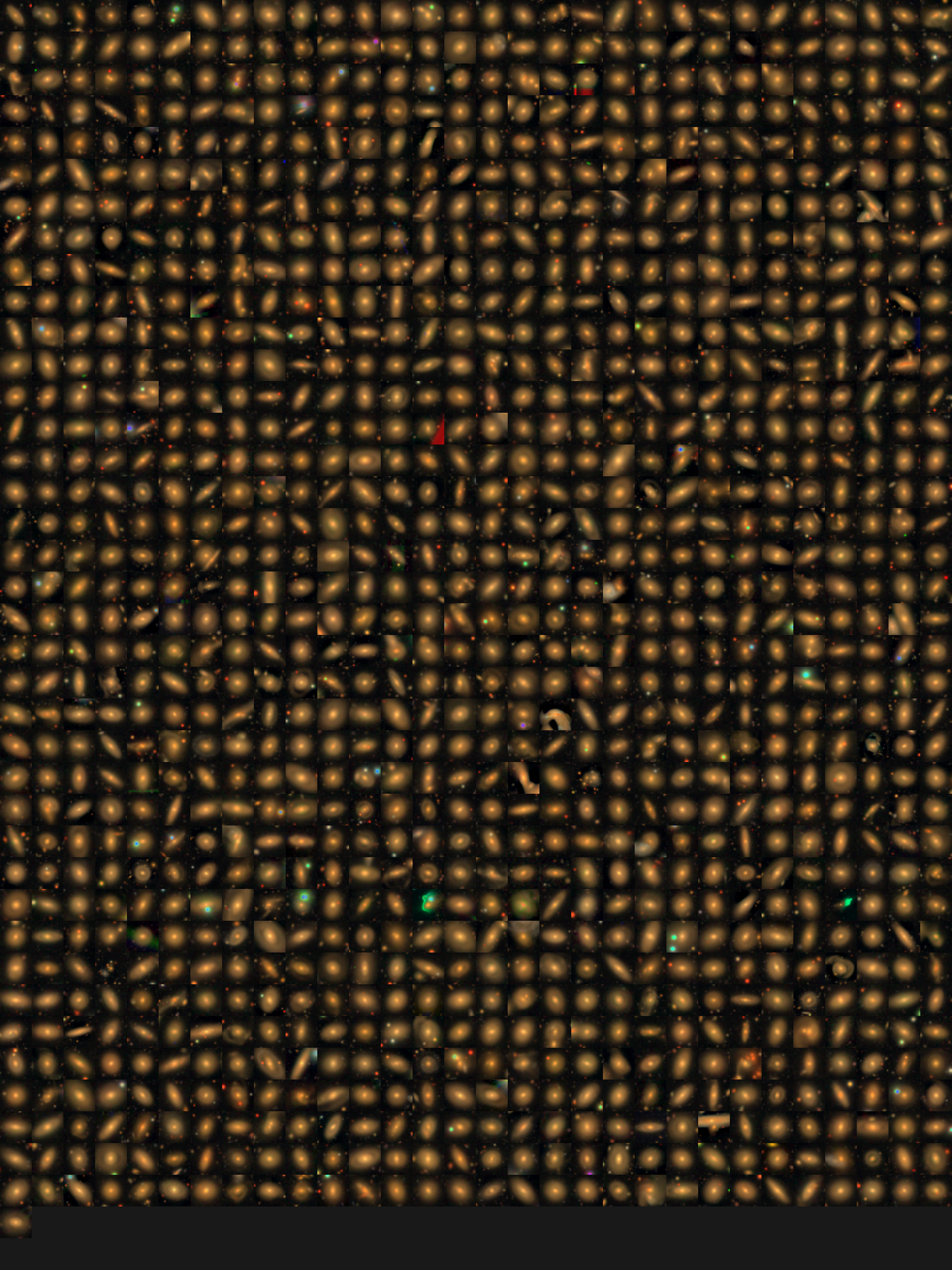}
\caption{
The secondary samples.
}
\label{fig2a}
\end{figure*}

\end{document}